\documentclass[12pt]{spieman}  
\usepackage[]{graphicx}
\usepackage{setspace}
\usepackage{tocloft}
\usepackage{amsmath}
\usepackage{multirow}
\title{Optimization of the multiple sampling and signal extraction in non-destructive exposures}

\author{Bogna Kubik\supscr{a}, Remi Barbier\supscr{a}, Alain Castera\supscr{a}, Eric Chabanat\supscr{a}, Sylvain Ferriol\supscr{a},\\  
Gerard Smadja\supscr{a}}

\affiliation{\supscrsm{a}Universit\'e de Lyon, Universit\'e Lyon 1, Lyon F-69003, France \\
CNRS/IN2P3, Institut de Physique Nucl\'eaire de Lyon, Villeurbanne F-69622, France
}

\cftpagenumbersoff{figure}
\cftpagenumbersoff{table} 
\begin{document} 
\maketitle 

\begin{abstract}
We derive the full covariance matrix formulae are derived for proper treatment of correlations in signal fitting procedures, extending the results from previous publications.

The straight line fits performed with these matrices demonstrate that a significantly higher signal to noise is obtained when the fluence exceeds 1 $e^-$/sec/pix in particular in long (several hundreds of seconds) spectroscopic exposures. The improvement arising from the covariance matrix is particularly strong for the initial intercept of the fit at t=0, a quantity which provides a useful redundancy to cross check the signal quality.
We demonstrate that the mode that maximizes the signal to noise ratio in all ranges of fluxes studied in this paper is the one that uses all the frames sampled during the exposure. 
While at low flux there is no restriction
on the organization of frames within groups for fluxes lower than 1 $e^-$/sec/pix, for fluxes exceeding this value the coadding of frames shell be avoided.
\end{abstract}


{\noindent \footnotesize{\bf Address all correspondence to}: Bogna Kubik, Institut de Physique Nucl\'eaire de Lyon, 4 rue Enrico Fermi, Villeurbanne F-69622, France Tel: +33 4 72 44 58 52; E-mail:  \linkable{bkubik@ipnl.in2p3.fr} }

\begin{spacing}{2}   

\section{Introduction}\label{sec:intro}
Up to now, many efforts have been made to minimize the readout noise of the integrating arrays in low-background experiments, such as spectroscopic imaging, where the readout noise of the array may remain the dominant source of noise even with integration time as long as of tens minutes. However, few studies have given attention to methods of signal fitting, as a simple least square fit provides a good signal estimate and an acceptable precision on the flux. The correlations in Poisson and readout noise that are intrinsic to the non-destructive readout of integrating arrays have been neglected, leading to a wrong estimation of signal noise. If a very stringent error budget is required it is crucial to take these correlations into account in the fitting procedure.

In ~\cite{Fowler1990} an algorithm of processing the array of pixels by averaging multiple reads was proposed to lower the readout noise as compared to the single read. The performances of Fowler sampling, correlated double sampling (CDS) and line fitting were studied in \cite{Garnett&Forrest1993}. The statement was made that both, Fowler sampling and line fitting are superior to the CDS in the readout noise limited performance. Moreover, the line fitting is superior to the optimal Fowler sampling by about 6\%. In the background limited regime, where the detector noise is dominated by the shot noise of the incident photons, the optimal signal to noise ratio is achieved in the CDS sampling. The variance of the integrated signal from data sampled continuously up the ramp was derived in \cite{Rauscher2007} and \cite{Vacca2004}. This noise model includes readout noise and shot noise on 
the integrated flux, taking into account the correlations between the multiple nondestructive reads.

Three main issues are addressed in this study. First we focus on the method of signal fits in different regimes of incident flux. We derive the formulae of covariance matrices for groups and group differences that take into account the correlations in readout noise and in Poisson noise arising from multiple reads during a single exposure. Next, we introduce these matrices into the fitting procedure and show that taking into account these correlations we can achieve a higher signal to noise ratio than with the line fitting. Finally, taking advantage of the possibility of nondestructive reads and the fit with the covariance matrix, we find out the optimal readout mode independently on the flux regime.

The equations for noise correlations in the MACC readout scheme are derived and compared to a simulated values in section \ref{sec:linear_fit}. The procedure giving the optimal signal to noise ratio are presented in section \ref{sec:best_fit} while the optimal readout parameters of the MACC acquisitions are found in section \ref{sec:best_mode}.

\newpage
\section{Linear fit of data with correlated errors}
\label{sec:linear_fit}
Infrared astronomy has been revolutionized by the introduction of two-dimensional array detectors, which have increased the efficiency of telescopes more than a thousandfold for direct imaging. The situation is quite different for low-background experiments, such as narrow band imaging or spectroscopy, where the readout noise of the array detectors may remain the dominant source of noise even in integrations of tens of minutes duration. Moreover, the possibility of non-destructive reads of the integrating pixels allowed to lower the readout noise by applying the procedure known as frames coadding (see \cite{Fowler1990}, \cite{Garnett&Forrest1993}, \cite{Rauscher2007} and \cite{Vacca2004} for definitions). 

It was already noticed in \cite{Anderson&Gordon2011} that the best calculation of slope and $y$-intercept must use all information including correlated and uncorrelated errors. A fit with the covariance matrix of equally sampled data was presented. The frames were sampled up the ramp and were not averaged into groups. It was shown that the fit with covariance matrix provides higher signal to noise ratio than a simple equally weighted least square fit. 

In this section, we derive the general expression of noise correlations between groups and group differences in the ramp sampled with the Multiple Accumulated readout scheme and construct the covariance matrix $C$ for line fitting and the matrix $D$ for fitting of differences. 

The values on the diagonal of $C$ and $D$ are the uncertainties of the
fitted points, whereas the off-diagonal elements are the covariances between
the different samples in the ramp. 

The diagonal entries are not only simple sums of the readout noise and shot noise, which would be the case for a simple up the ramp sampling without coadding, but include the correlations in the shot noise after the coadding procedure.
Moreover, if the data points were uncorrelated, then the off-diagonal matrix elements would be all zero. However, we do have correlated data in addition to correlations resulting from the coadding procedure: the shot noise is correlated in the line fitting while the readout noise correlates points in the difference fit. 

The real power of using matrix notation arises from the covariance matrix, as it includes information about the degree of correlation between samples. Furthermore, the slope and $y$-intercept uncertainties resulting from using this covariance matrix automatically take into account the correlated and uncorrelated errors, provided they are included in the covariance matrix. Another benefit of using matrix notation with a covariance matrix is that we can add other noise terms easily by calculating the appropriate covariance matrix - this would be the case of the $1/f$ noise which correlates all the groups in the ramp and should be incorporated into the covariance matrix to fit real data. It is however out of the scope of this paper and we leave this subject for later investigations.

\subsection{UTR and MACC readout modes}
\begin{figure}[ht!]
    \begin{center}
        \begin{tabular}{cc}
            \includegraphics[scale=0.35]{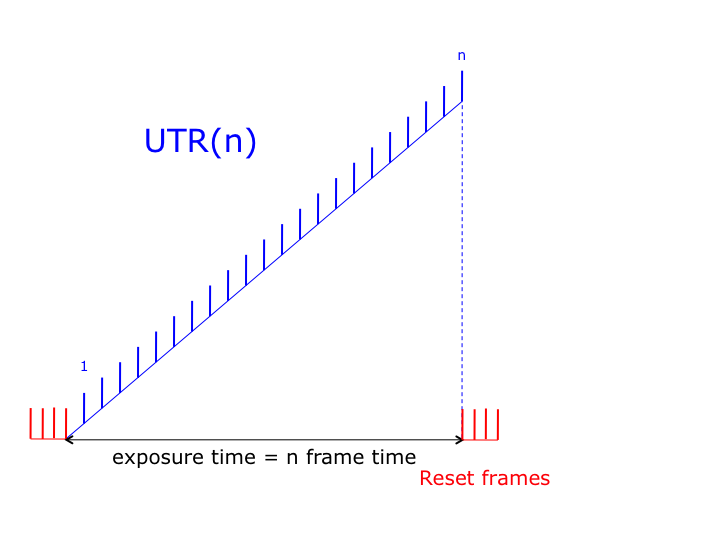}
            \includegraphics[scale=0.35]{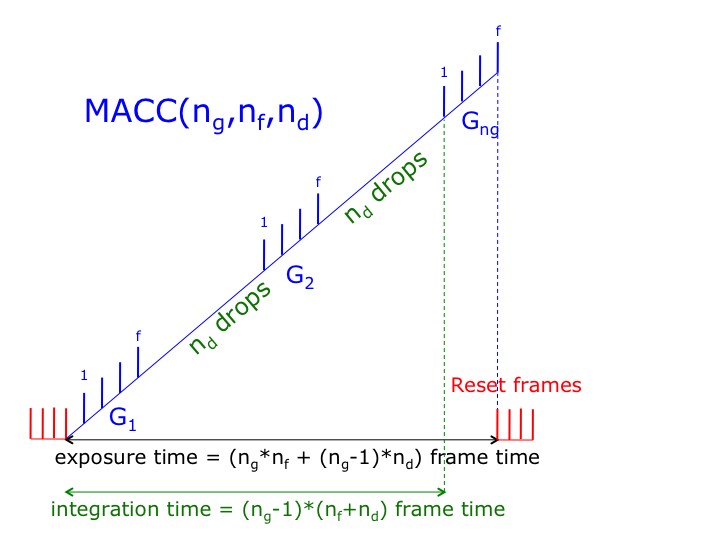}
        \end{tabular}
    \end{center}
    \caption[Schematic representation of readout modes]{\label{fig:readout_modes}Left: up the ramp (UTR) sampled data at a constant frame cadence. A frame is the unit of data that results
from sequentially clocking through and reading out a rectangular
area of pixels. Right: Multiple Accumulated sampling MACC($n_g$, $n_f$, $n_d$) with $n_g$ - number of equally spaced groups sampled up the ramp, $n_f$ - number of
averaged frames per group and $n_d$ - number of dropped frames between two successive groups.}
\end{figure}

For most science observations, NIR detectors acquire
up the ramp (UTR) sampled data at a constant frame cadence as presented in the left panel of Figure \ref{fig:readout_modes}. A frame is the unit of data that results
from sequentially clocking through and reading out a rectangular
area of pixels. As in the UTR mode frames suffer from a high readout noise it is common to average the frames within groups. We name this readout pattern a Multiple Accumulated (MACC) sampling and frequently we use the abbreviation MACC($n_g$, $n_f$, $n_d$) where $n_g$ is the number of equally
spaced groups sampled up the ramp, $n_f$ is the number of
averaged frames per group and $n_d$ is the number of dropped frames between two successive groups. (see right panel of Figure \ref{fig:readout_modes} for a schematic representation of MACC readout mode.)

One advantage of up the ramp sampled data for space platforms
is that cosmic rays can potentially be rejected with minimal
data loss. 
The total exposure time is the time between digitizing pixel [0, 0] in the first frame of the first group
and digitizing the same pixel in the lats frame of the last group.

\subsection{Covariance matrix for groups}
\label{subsec:covariance_matrix_groups}
In this section, we derive the general expression of noise correlations between groups in the ramp sampled with the MACC readout scheme. The noise model for the MACC readout fitted by a straight line was derived in \cite{Rauscher2007} using the error propagation formalism. We agree with the total variance presented in this paper, however we notice that the covariance matrix terms are not properly organized, which would potentially lead to a wrong signal determination if the total covariance matrix would be used. It would also be more difficult to add
other noise terms to the Equation (1) in \cite{Rauscher2007} which, like our method, only include readout noise and photon shot noise. We propose here an alternative way to derive the covariance matrix elements and show that using this matrix in the fitting procedure we achieve a higher signal to noise ratios than in the simple line fitting by a least square fit. 

We define $f_0$ as the signal accumulated between the last reset and the first readout of the pixel\footnote{In what follows we assume that the array is reset pixel by pixel so that the average signal between the last reset and the first readout is equal to the average signals between successive frames in the ramp.}. $f_i^{(k)}$ is a signal accumulated between frame $S^{(k)}_i$ and $S_{i+1}^{(k)}$ in group $k$ (first to first pixel integration time). $d^{(k)}$ is the signal accumulated between the last frame of group $G_k$ and the first frame of group $G_{k+1}$. $\rho_i^{(k)}$ is the readout noise of frame $S^{(k)}_i$ in group $k$. The inter-frame integrated fluxes are stochastically independent and follow the Poisson distribution with the mean value $f$. The same holds for the inter group integrated fluxes which have the mean $d$. The readout noise, assumed here uncorrelated, follow Gaussian distribution with the width $\sigma_R$
\begin{align}
    \langle \delta f_i^{(k)} \delta f_j^{(l)} \rangle & = f\,\delta_{kl}\delta_{ij}\, , \nonumber \\
    \langle \delta d^{(k)} \delta d^{(l)} \rangle & = d\,\delta_{kl}\, , \\
    \langle \delta \rho_i^{(k)} \delta \rho_j^{(l)} \rangle & = \sigma_R\, \delta_{kl}\delta_{ij}\, . \nonumber
\end{align}
We assume that in a single exposure we have $n$ groups of $m$ coadded frames. The reasoning presented below should be done for each pixel independently.
The consecutive frames up the ramp $S_i^{(1)}$ sampled within the first group $G_1$ are equal to
\begin{align*}
        S_1^{(1)} & =  \rho_1^{(1)} + f_0\, , \\
        S_2^{(1)} & =  \rho_2^{(1)} + f_0 + f_1^{(1)}\, , \\
        S_3^{(1)} & =  \rho_3^{(1)} + f_0 + f_1^{(1)} + f_2^{(1)}\, , \\
        \dots \\
        S_m^{(1)} & =  \rho_m^{(1)} + f_0 + \dots + f_{m-1}^{(1)}\, .~~~~~~~~~~~~~~~~~~~~~~~~~~~~~~~~~
\end{align*}
In the second group we have:
\begin{align*}
        S_1^{(2)} & =  \rho_1^{(2)} + f_0 + \dots + f_{m-1}^{(1)} + d^{(1)}\, , \\
        S_2^{(2)} & =  \rho_2^{(2)} + f_0 + \dots + f_{m-1}^{(1)} + d^{(1)} + f_1^{(2)}\, , \\
        S_3^{(2)} & =  \rho_3^{(2)} + f_0 + \dots + f_{m-1}^{(1)} + d^{(1)} + f_1^{(2)} + f_2^{(2)}\, , \\
        \dots \\ 
        S_m^{(2)} & =  \rho_m^{(2)} + f_0 + \dots + f_{m-1}^{(1)} + d^{(1)} + f_1^{(2)} + \dots + f_{m-1}^{(2)}\, .
\end{align*}
so that the coadded group equals
\begin{align*}
    G_1 & = \frac{1}{m}\sum_{i=1}^m S_i^{(1)} = \frac{1}{m}\sum_{i=1}^m\rho_i^{(1)} + \frac{1}{m}\sum_{i=1}^{m-1}(m-i)f_i^{(1)} + f_0\, , \\
    G_2 & = \frac{1}{m}\sum_{i=1}^m S_i^{(2)} = \frac{1}{m}\sum_{i=1}^m\rho_i^{(2)} + \frac{1}{m}\sum_{i=1}^{m-1}(m-i)f_i^{(2)} + f_0 + \sum_{i=1}^{m-1}f_i^{(1)} + d^{(1)} \,.~~~~~~~~~~
\end{align*}
Those equations can be easily generalized to a group $G_k$:
\begin{align}
    G_k & = \frac{1}{m}\sum_{i=1}^m S_i^{(k)} = \frac{1}{m}\sum_{i=1}^m\rho_i^{(k)} + \frac{1}{m}\sum_{i=1}^{m-1}(m-i)f_i^{(k)} + f_0 + \sum_{j=1}^{k-1}\left(\sum_{i=1}^{m-1}f_i^{(j)} + d^{(j)}\right)\ .
\end{align}
Using the independence of the stochastic fluctuations we can easily write the variance of a group $G_{k}$:
\begin{equation}\label{eq_Ckk}
    C_{kk} \equiv \langle ( \delta G_k )^2 \rangle = \frac{\sigma_R^2}{m} + (k-1)(m-1)f + (k-1)d + \frac{(m+1)(2m+1)f}{6m}\, ,
\end{equation}
where the last term comes form the correlations of Poisson noise between frames in the same group.
The correlations between two different groups can be also easily computed. For $k<l$ we find
\begin{eqnarray}\label{eq_Ckl}
    C_{kl} \equiv \langle \delta G_k \delta G_l \rangle = (k-1)(m-1)f + (k-1)d + \frac{1}{2}(m+1)f\, ,
\end{eqnarray}
where the last term arises from the correlations of Poisson noise between the groups $k$ and $l$. 

As an example we give the covariance matrix elements calculated using Equations (\ref{eq_Ckk}) and (\ref{eq_Ckl}) in one of the planned Euclid science readout modes MACC(4,16,0) with the exposure time of about 100 sec, for a flux of $2.5$ $e^{-}$/sec and a frame readout noise $\sigma_R = 10$ $e^-$. The theoretical prediction is confirmed in our Monte-Carlo simulations where the covariance matrix is calculated as $C_{kl} = \langle G_k G_l \rangle - \langle G_k \rangle ~\langle G_l \rangle$ over 10000 simulated ramps:
\begin{displaymath}
C_{kl}^{\textrm{simu}} = \left(
    \begin{tabular}{cccc}
   28.47 &  32.01 &   32.39 &  32.33 \\
   32.01 &  86.87 &   91.00 &  89.89 \\
   32.39 &  91.00 &  146.34 &  148.75 \\
   32.33 &  89.89 &  148.75 &  204.57 
    \end{tabular}
    \right) ~~~~
    C_{kl} = \left(
    \begin{tabular}{cccc}
    28.16 &  31.88 &   31.88 &   31.88 \\
    31.88 &  88.16 &   91.88 &   91.88 \\
    31.88 &  91.88 &  148.16 &  151.88 \\
    31.88 &  91.88 &  151.88 &  208.16
    \end{tabular}
    \right) \, .
\end{displaymath}
\subsection{Covariance matrix for group differences}
\label{subsec:covariance_matrix_differences}
In the fit of differences of averaged frames the shot noise correlations resulting from the coadding procedure are present and additionally the readout noise correlates different sampling points.
Another advantage of taking the group differences instead of individual group values to compute the flux is that we are not sensitive to the reset noise and that a large part of correlations are removed.
With the same notations as in the previous section we can derive the covariance matrix for group differences. For the first two groups we have:
\begin{align*}
        S_1^{(2)} - S_1^{(1)} & = f_1^{(1)} + f_2^{(1)} + f_3^{(1)} + \dots + f_{m-1}^{(1)} + d^{(1)} + \rho_1^{(2)} - \rho_1^{(1)} \\
        S_2^{(2)} - S_2^{(1)} & = ~~~~~~~~\,f_2^{(1)} + f_3^{(1)} + \dots + f_{m-1}^{(1)} + d^{(1)} + \rho_2^{(2)} - \rho_2^{(1)} + f_1^{(2)}\\
        S_3^{(2)} - S_3^{(1)} & = ~~~~~~~~\,~~~~~~~~\,f_3^{(1)} + \dots + f_{m-1}^{(1)} + d^{(1)} + \rho_3^{(2)} - \rho_3^{(1)} + f_1^{(2)} + f_2^{(2)}\\
        \dots\\
        S_m^{(2)} - S_m^{(1)} & = ~~~~~~~~\,~~~~~~~~\,~~~~~~~~\,~~~~~~~~\,~~~~~~~~\,\,d^{(1)} + \rho_m^{(2)} - \rho_m^{(1)} + f_1^{(2)} + f_2^{(2)} + \dots + f_{m-1}^{(2)}
\end{align*}
and the coadded group difference is
\begin{equation*}
    G_2 - G_1 = \frac{1}{m}\sum_{i=1}^{m}\left( S_i^{(2)} -  S_i^{(1)} \right) = d^{(1)} + \frac{1}{m} \sum_{i=1}^{m-1}(if_i^{(1)} + (m-i)f_i^{(2)}) + \frac{1}{m}\sum_{i=1}^m(\rho_i^{(2)} - \rho_i^{(1)})\, ,
\end{equation*}
which generalizes easily to
\begin{equation}
    G_{k+1} - G_k = d^{(k)} + \frac{1}{m} \sum_{i=1}^{m-1}(if_i^{(k)} + (m-i)f_i^{(k+1)}) + \frac{1}{m}\sum_{i=1}^m(\rho_i^{(k+1)} - \rho_i^{(k)})\, .~~~~~~~~~~~~~~~~~~~~~
\end{equation}
The diagonal terms in the covariance matrix will be then all equal to the same value:
\begin{equation}
   D_{kk}\equiv \langle ( \delta(G_k - G_{k-1}) )^2 \rangle = \frac{2\sigma_R}{m} + d + \frac{(m-1)(2m-1)}{3m}f\, .
\end{equation}
The off-diagonal terms are correlated by the Poisson noise and the readout noise:
\begin{align*}
\langle \delta(G_2 - G_1) \delta(G_3 - G_2) \rangle = ~& \langle
  \left( \delta d^{(1)} + \frac{1}{m} \sum_{i=1}^{m-1}(i\delta f_i^{(1)} + (m-i)\delta f_i^{(2)}) + \frac{1}{m}\sum_{i=1}^m(\delta\rho_i^{(2)} - \delta\rho_i^{(1)}) \right) \nonumber \\
&\left( \delta d^{(2)} + \frac{1}{m} \sum_{i=1}^{m-1}(i\delta f_i^{(2)} + (m-i)\delta f_i^{(3)}) + \frac{1}{m}\sum_{i=1}^m(\delta\rho_i^{(3)} - \delta\rho_i^{(2)}) \right) \rangle \nonumber \\
= ~& \langle  \left( \frac{1}{m} \sum_{i=1}^{m-1}(m-i)\delta f_i^{(2)} + \frac{1}{m}\sum_{i=1}^m\delta\rho_i^{(2)}  \right) \\
            &\left( \frac{1}{m} \sum_{i=1}^{m-1} i\delta f_i^{(2)}    - \frac{1}{m}\sum_{i=1}^m\delta\rho_i^{(2)}  \right)\rangle \nonumber\\
= ~& \frac{(m^2-1)}{6m}f-\frac{\sigma_R^2}{m} \nonumber \, .
\end{align*}
Generalizing this formula we find
\begin{equation}
    D_{kl} \equiv \langle \delta(G_{k+1} - G_k) \delta(G_{l+1} - G_l) \rangle = \frac{(m^2-1)}{6m}f-\frac{\sigma_R^2}{m\, ,}
\end{equation}
all other terms of the matrix are equal to zero if we take into account only shot and readout noise contributions. For the same simulated values as in the previous section: MACC(4,16,0) readout mode, an incident flux of $2.5$ $e^{-}$/sec and a frame readout noise $\sigma_R = 10$ $e^-$ we obtain the following matrix elements confirmed by the simulations:
\begin{displaymath}
D_{kl}^{\textrm{simu}} = \left(
    \begin{tabular}{ccc}
   43.92 &  1.54 & 0.14 \\
    1.54 & 43.62 & 2.85 \\
    0.14 &  2.85 & 45.80 \\
    \end{tabular}
    \right) ~~~~
    D_{kl} = \left(
    \begin{tabular}{ccc}
   44.56 &  1.72 &  0    \\
    1.72 & 44.56 &  1.72 \\
       0 &  1.72 & 44.56 \\
    \end{tabular}
    \right)
\end{displaymath}
We can see that the impact of the covariance matrix will be much weaker in this case than in the fit of groups.

\newpage
\section{Optimal fitting method of coadded data}
\label{sec:best_fit}
To demonstrate how well this calculation of the uncertainties fits simulated data we take as example of typical photometric and spectroscopic exposures and compare the signal to noise ratio (SNR) reached in four methods of flux fitting:
\begin{itemize}
    \item a simple unweighted least square fit (LSF)
    \item a least square fit with the full covariance matrix for coadded groups (COV)
    \item an unweighted least square fit of group differences (LSFd)
    \item a least square fit with the covariance matrix for group differences (COVd)
\end{itemize}
For a refresher on calculating a linear fit using matrices as well as the use of the covariance matrix, see \cite{Hogg2010}. For each pixel, data are usually well modeled
by a two-parameter line of the form
\begin{equation}
    s = a t + b\, ,
\end{equation}
where $s$ if the integrated signal in units of $e^-$, $a$ is the slope, $b$ is the $y$-intercept and $t$ is time.
We show the impact of using the full covariance matrix on the signal fitting in Tables (\ref{tab:spectro_exposure_fits}) and (\ref{tab:spectro_exposure_intercept_noise}) in an example of spectroscopic exposure of 549 sec and a photometric exposure of 96 sec total exposure time. As the reference readout modes we take the proposed science readout modes for the NISP instrument in the Euclid mission: a MACC mode with 15 groups of 16 coadded frames and 9 drops MACC(15,16,9) in spectroscopic exposures and a MACC(4,16,0) in photometric exposures. 

\begin{figure}[ht!]
    \begin{center}
        \begin{tabular}{cc}
            \includegraphics[scale=0.4]{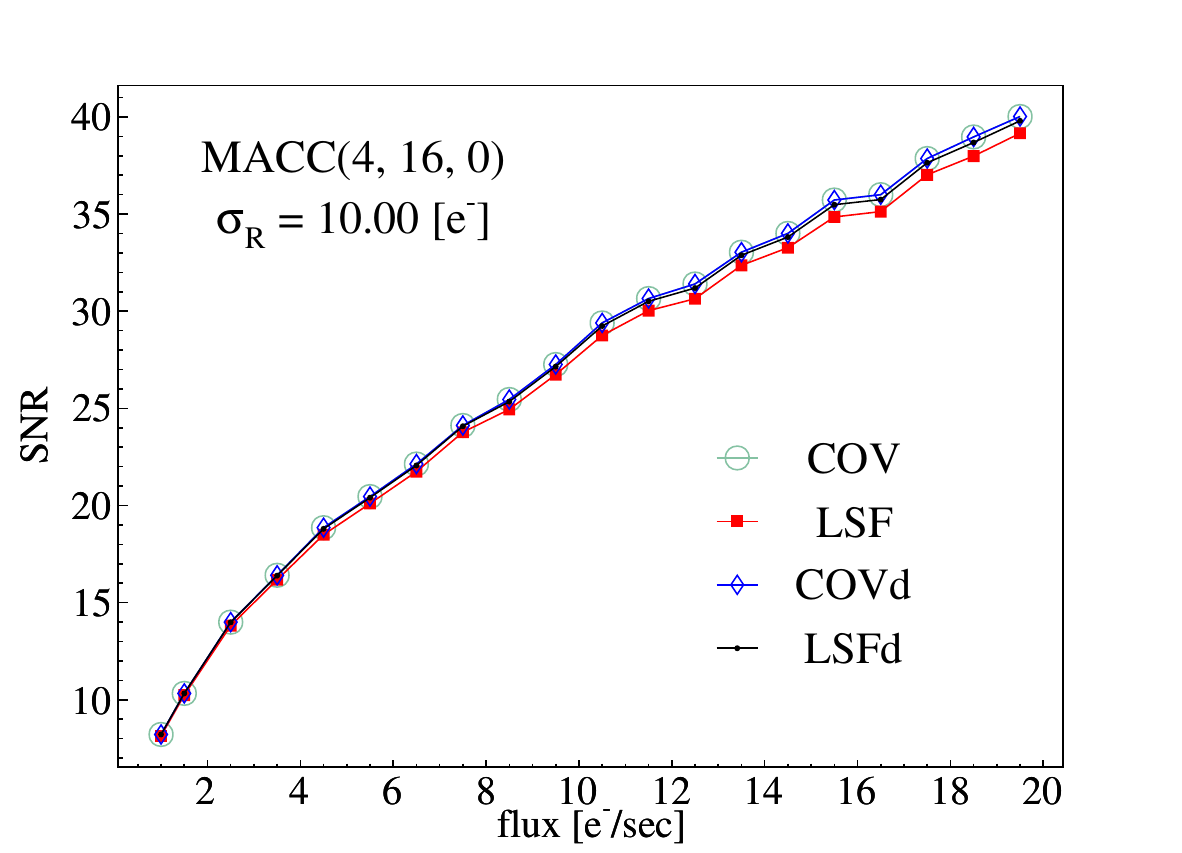}
            \includegraphics[scale=0.4]{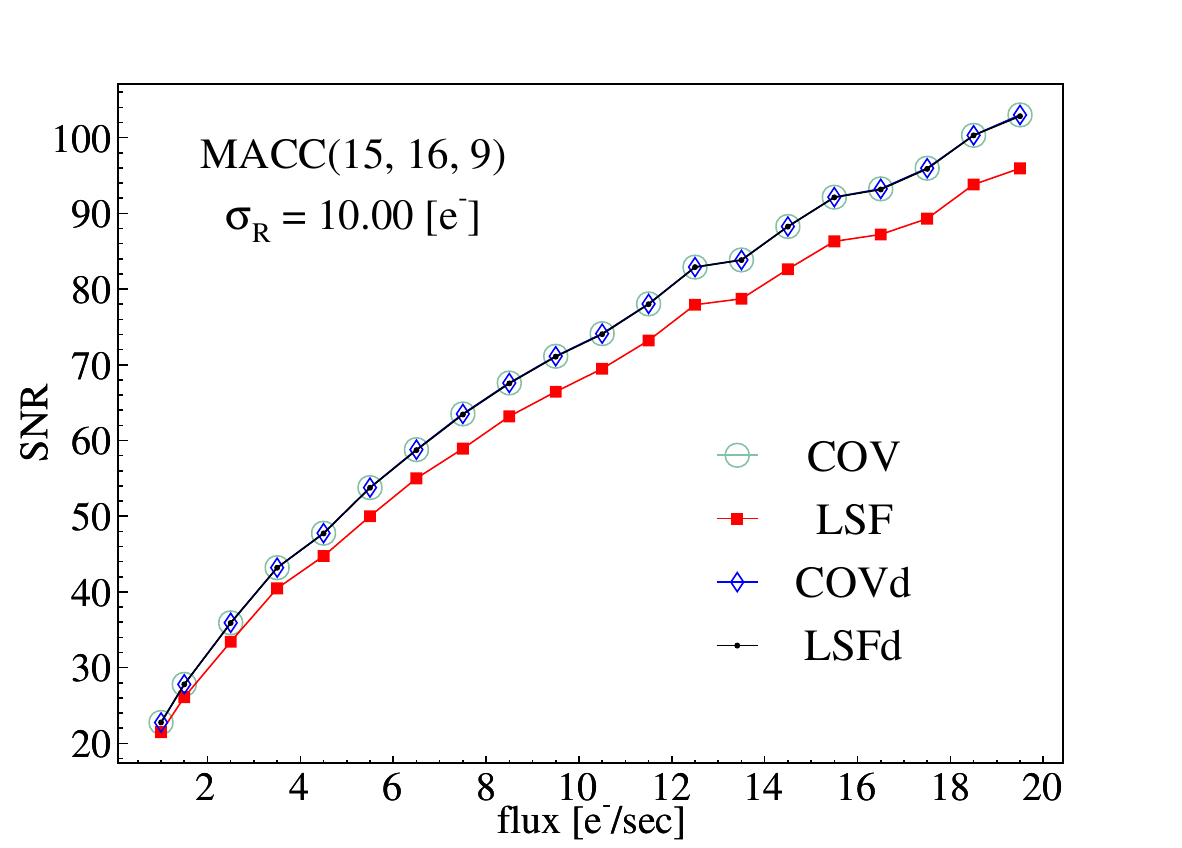}\\
        \end{tabular}
    \end{center}
    \caption[Signal to noise ratio for different fitting methods in photometric and spectroscopic exposures]{\label{fig:SNR_ratio}Signal to noise ratio and $y$-intercept noise (in [$e^-$]) using the fits described in section \ref{sec:best_fit}. The simulated ramps were fitted by four different methods and the $y$-intercept and slope standard deviations were computed. The fits with covariance matrices and the simple fit of differences give signal to noise ratios similar but higher than the LSF fit.}
\end{figure}

\begin{figure}[ht!]
    \begin{center}
        \begin{tabular}{cc}
            \includegraphics[scale=0.4]{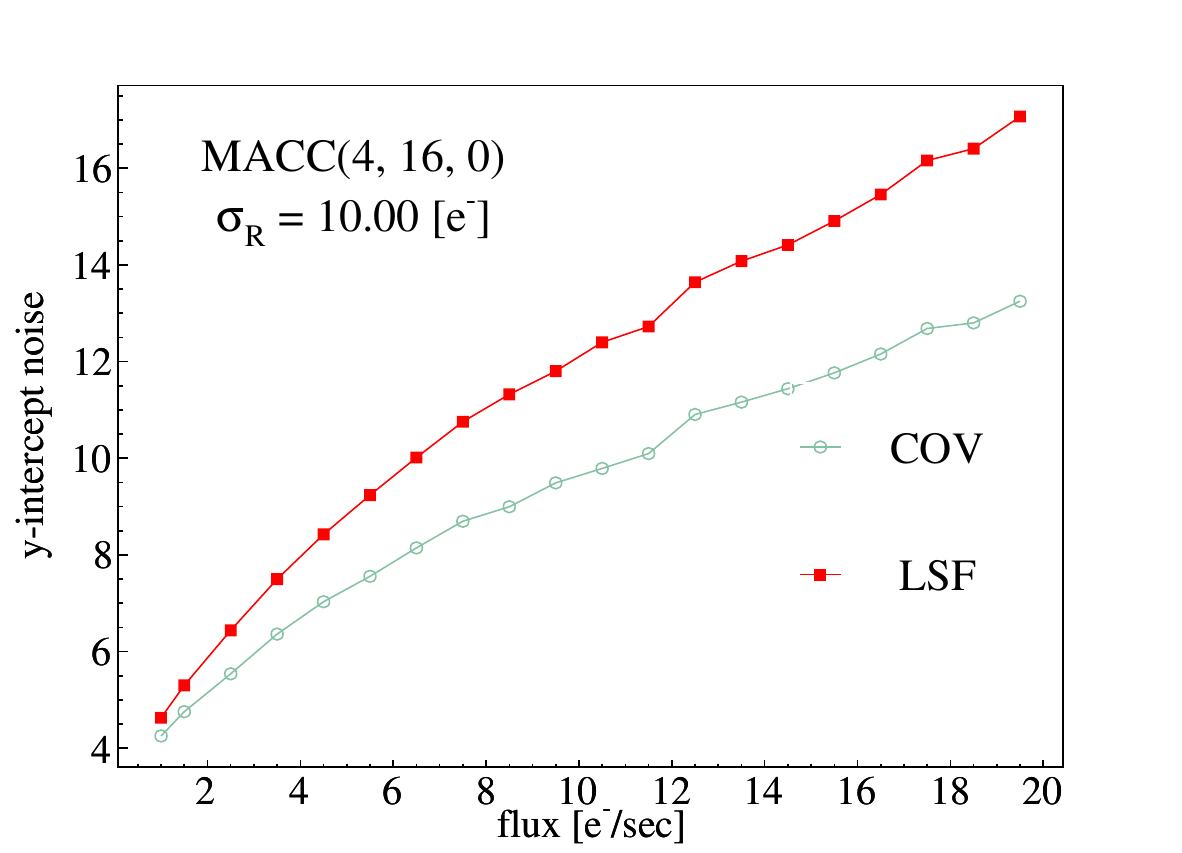}
            \includegraphics[scale=0.4]{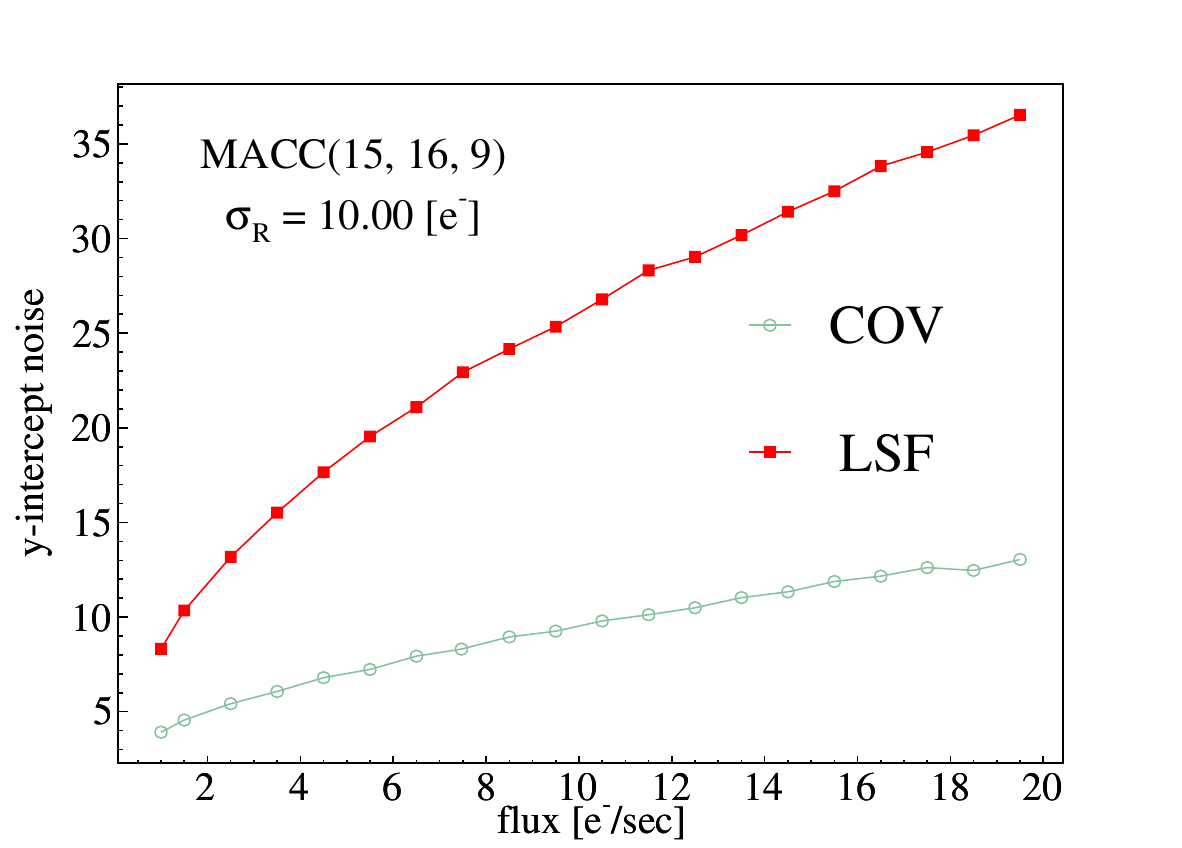}
        \end{tabular}
    \end{center}
    \caption[$y$-intercept noise for different fitting methods in photometric and spectroscopic exposures]{\label{fig:yintercept_noise}$y$-intercept noise (in [$e^-$]) using the fits described in section \ref{sec:best_fit}. The simulated ramps were fitted by four different methods and the $y$-intercept and slope standard deviations were computed. The benefit from the COV fit is that we have the information about the $y$-intercept which could be important in cosmic ray rejection algorithms and in persistence or nonlinearity correction.}
\end{figure}

In Figures (\ref{fig:SNR_ratio}) and (\ref{fig:yintercept_noise}) we show the signal to noise ratio and the $y$-intercept noise computed as standard deviation over 10,000 simulated ramps, including readout noise $\sigma_R = 10~e^-$/sec and shot noise, for a wide range of fluxes. We find out that the COV, COVd and LSFd fits give approximately the same signal to noise ratio which is systematically higher of about 5-7\% (dependent on the flux value) than the LSF fit in the spectroscopic exposures. For shorter exposure times, as the photometric exposures of about 100 sec the benefit of using covariance matrices or fit of differences is of about 1\% on the SNR. The fit of differences LSFd has the same performances that the fits with covariance matrices reaching higher SNR than the simple LSF of groups. 

While in the fits of differences the $y$-intercept is not defined, we see the huge impact on the $b$ parameter noise in COV fit which lowers the noise by 50\% (60\%) in spectroscopic exposures for 1 (10) $e^-$/sec fluxes respectively. In the photometric exposures the $y$-intercept noise is lowered by 6\% to 20\% depending on the flux value. The information about the $y$-intercept is crucial in cosmic ray rejection algorithms and in persistence or nonlinearity corrections.
\renewcommand{\arraystretch}{1.5}
\begin{table}[ht!]
{\footnotesize
    \begin{center}
        \begin{tabular}{ccc|cccc|cccc}
        \hline
         \multicolumn{3}{r}{}                                & \multicolumn{4}{c}{\bf SNR in spectroscopic exposure}      & \multicolumn{4}{c}{\bf SNR in photometric exposure} \\
        \hline
        \multicolumn{3}{r}{\it method}                       & {\it LSF} & {\it COV} & {\it LSFd} & {\it COVd}            & {\it LSF} & {\it COV} & {\it LSFd} & {\it COVd} \\
        \hline
        \multirow{2}{*}{\rotatebox{90}{\it flux}}  & \multirow{2}{*}{\rotatebox{90}{\it $e^-$/sec}}  
                                                  & {\it 1}  & 21.49     & 22.68     &   22.68    &  22.68    &  8.16     & 8.24      & 8.24       & 8.24       \\
                                                  & & {\it 10} & 68.42     & 73.05     &   73.03    &  73.05    & 27.84     & 28.35     & 28.25      & 28.35      \\
        \hline
        \end{tabular}
         \caption[Signal to noise ratios in spectroscopic and photometric exposures]{\label{tab:spectro_exposure_fits} Signal to noise ratios in spectroscopic MACC(15,16,9) ($t_{exp} = 549$ sec) and photometric MACC(4,16,0) ($t_{exp} = 96$ sec) exposures for different methods of signal fitting.}
    \end{center}
    }
\end{table}

\begin{table}[ht!]
{\footnotesize
    \begin{center}
        \begin{tabular}{ccc|cc|cc}
        \hline
         \multicolumn{3}{r}{}                                & \multicolumn{4}{c}{\bf $y$-intercept noise}                    \\
         \multicolumn{3}{r}{}                                & \multicolumn{2}{c}{\bf spectroscopic exposure} & \multicolumn{2}{c}{\bf photometric exposure} \\
        \hline
        \multicolumn{3}{r}{\it method}                       & {\it LSF} & {\it COV}               & {\it LSF} & {\it COV} \\
        \hline
        \multirow{2}{*}{\rotatebox{90}{\it flux}}  & \multirow{2}{*}{\rotatebox{90}{\it $e^-$/sec}}  
                                                   &{\it 1}  & 8.32      & 3.92                   & 4.47      & 4.19 \\
                                                 & &{\it 10} & 26.79     & 9.80                   & 12.31      & 9.91 \\
        \hline
        \end{tabular}
         \caption[$y$-intercept noise in spectroscopic and photometric exposures]{\label{tab:spectro_exposure_intercept_noise} $y$-intercept noise in spectroscopic MACC(15,16,9) ($t_{exp} = 549$ sec) and photometric MACC(4,16,0) ($t_{exp} = 96$ sec) exposures for different methods of signal fitting.}
    \end{center}
    }
\end{table}

\newpage
\section{Optimal readout mode for a given integration time}
\label{sec:best_mode}
For most science observations the Euclid-NISP detectors will acquire up the ramp sampled data at a constant cadence of one frame every $\approx$1.3 s which will be then averaged into $n_g$ equally spaced groups with $n_f$ frames per group. One advantage of up the ramp sampled data for detector systems of a space telescope is that cosmic rays can potentially be rejected or corrected with minimal data loss. We do not address here the issue of the efficiency of cosmic rays rejection but, following the conclusions of \cite{Anderson&Gordon2011}, we assume that the maximal number of samples increases the efficiency of cosmic rays detection. It is however not obvious whether the maximal number of samples (i.e.. groups of coadded frames) maximizes the signal to noise ratio assuming absence of cosmic ray or a possibility of correction of its impact. We focus here on the optimal arrangement of groups and frames along the ramp in order to maximize the SNR ratio.

\begin{figure}[ht!]
    \begin{center}
        \begin{tabular}{cc}
            \includegraphics[scale=0.4]{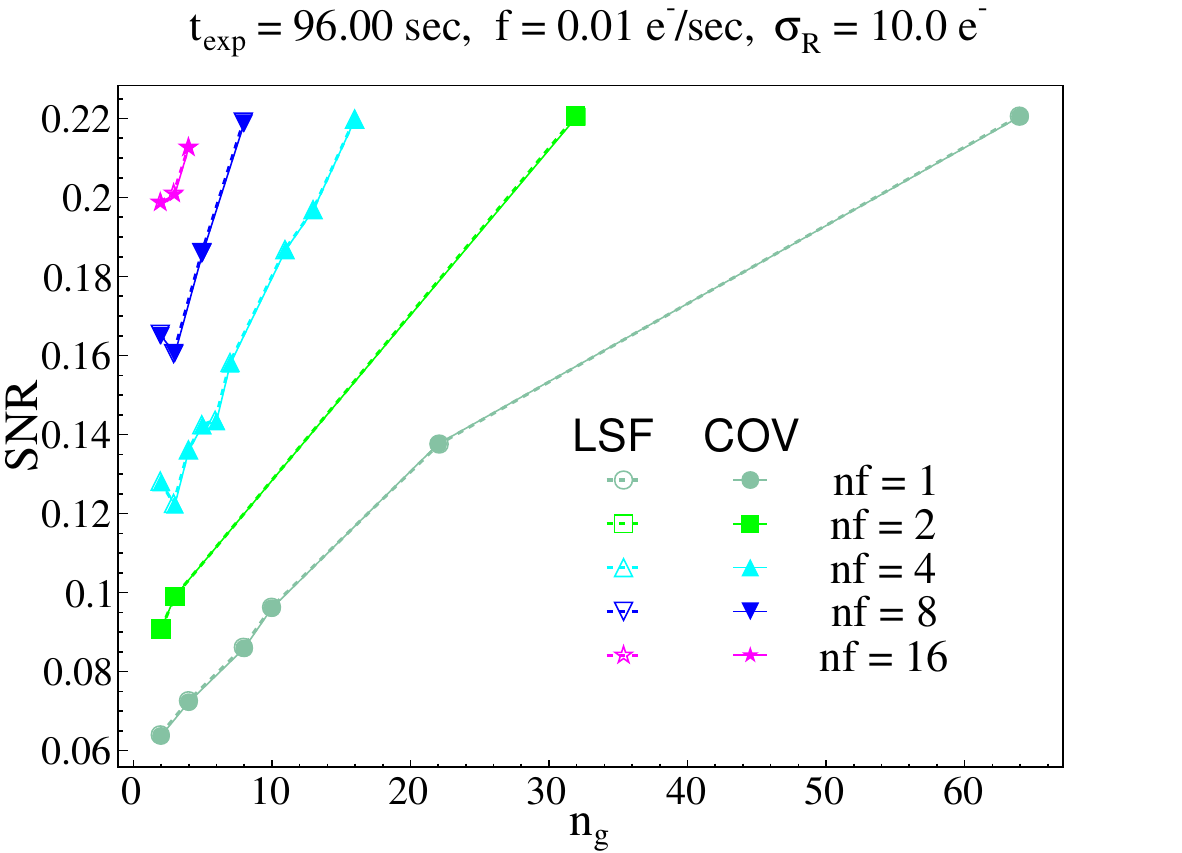}
            \includegraphics[scale=0.4]{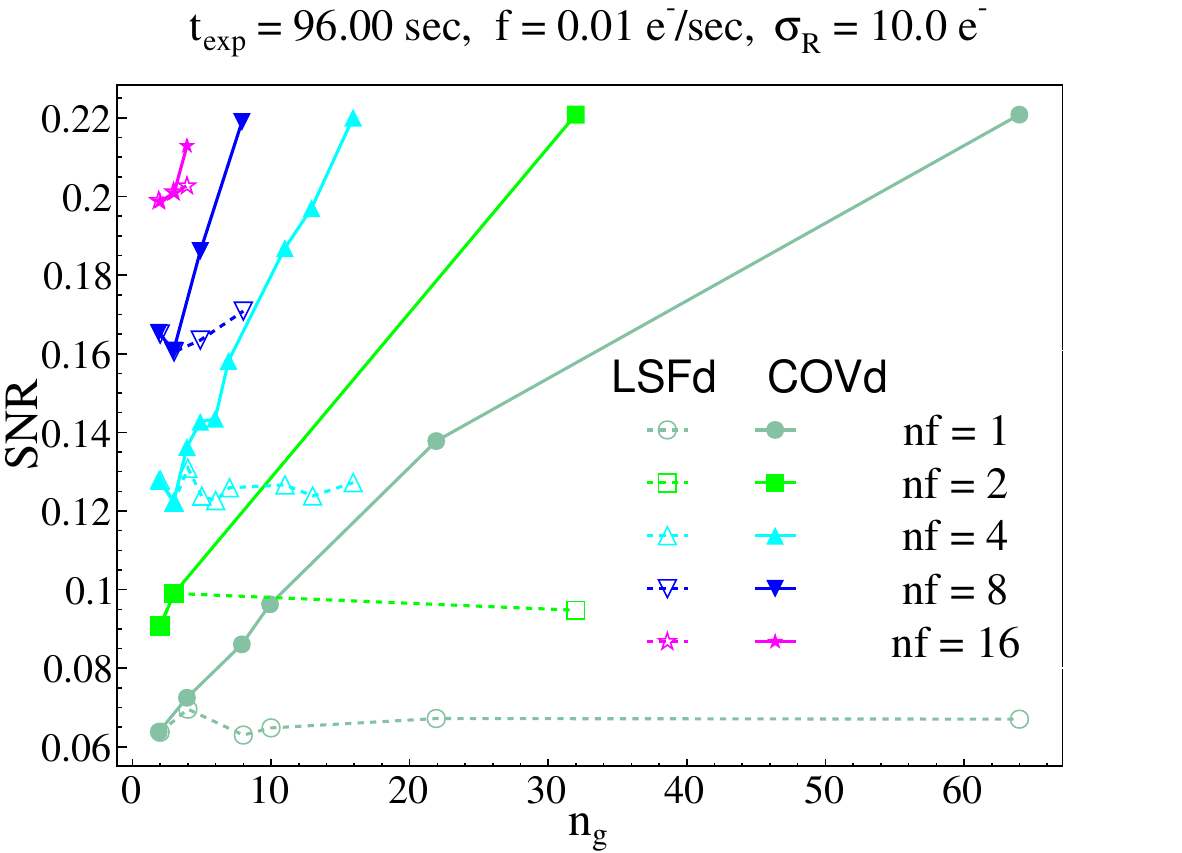}\\
            \includegraphics[scale=0.4]{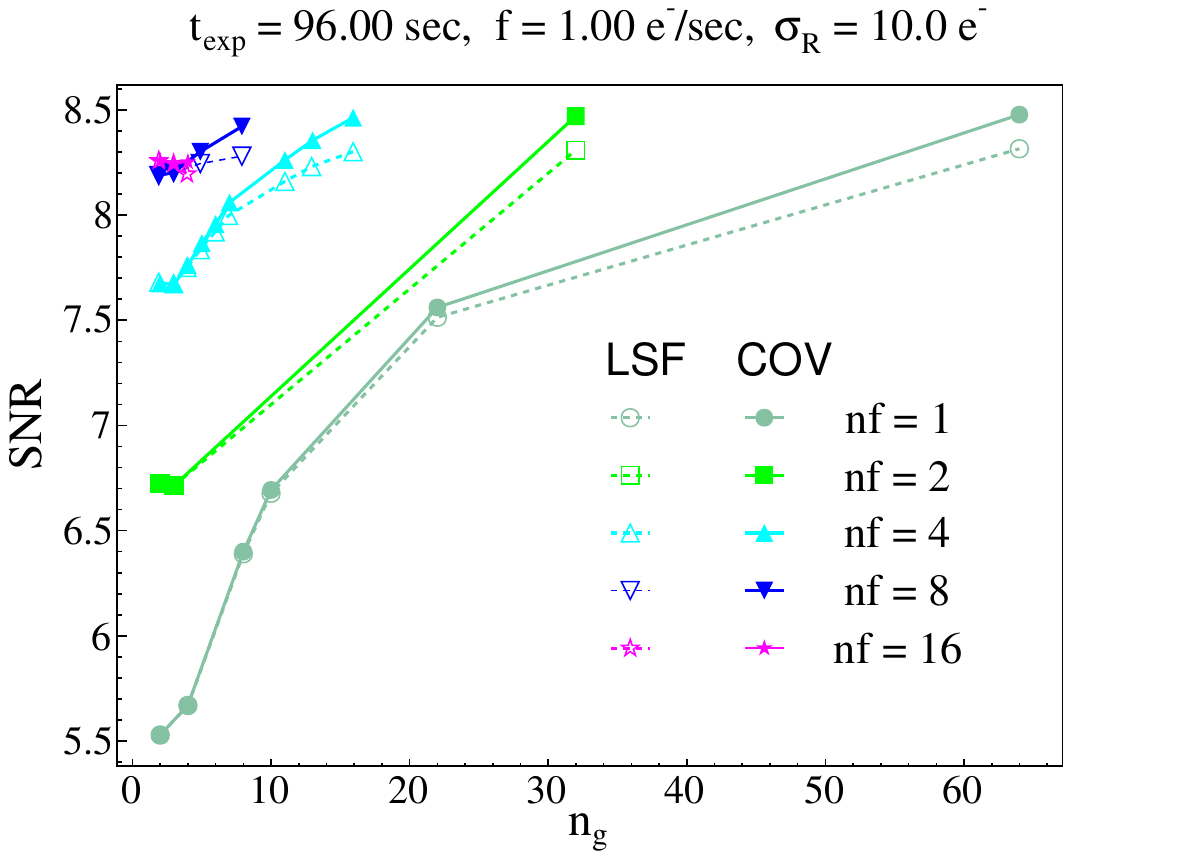}
            \includegraphics[scale=0.4]{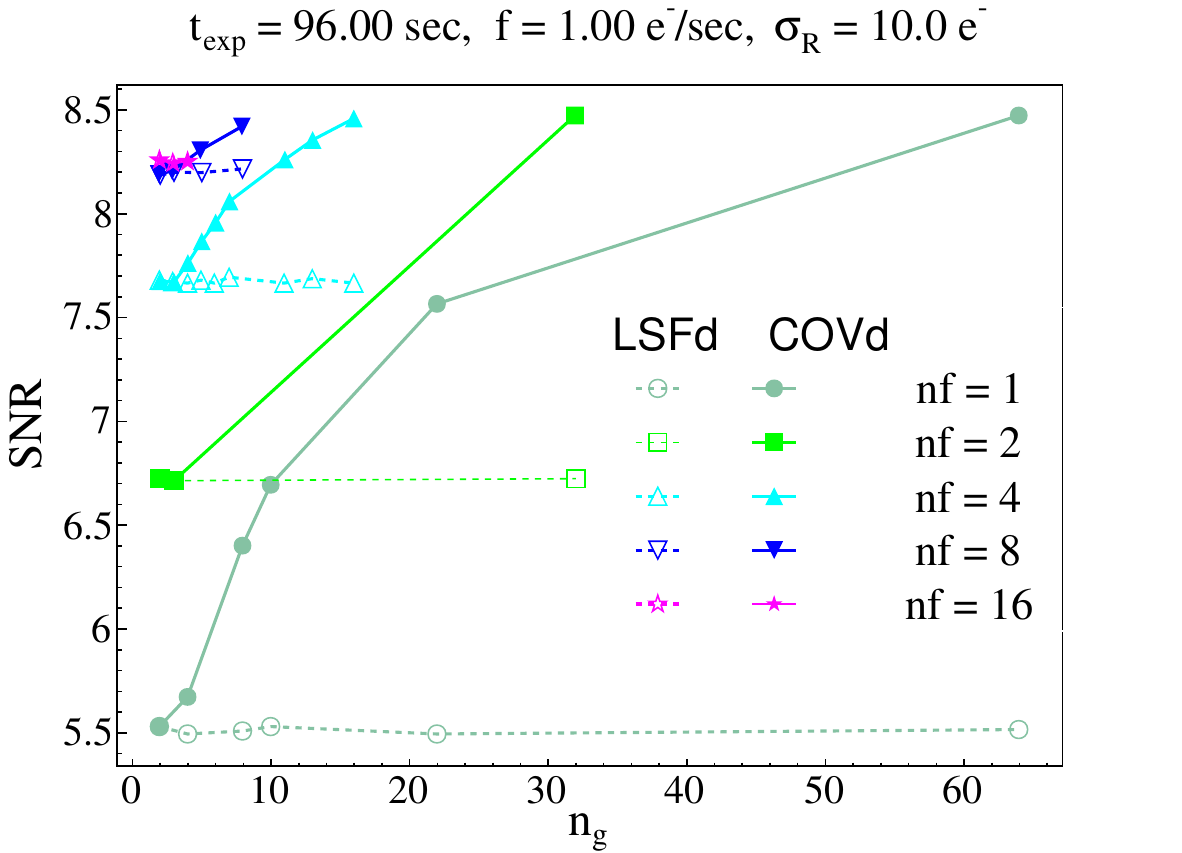}\\
            \includegraphics[scale=0.4]{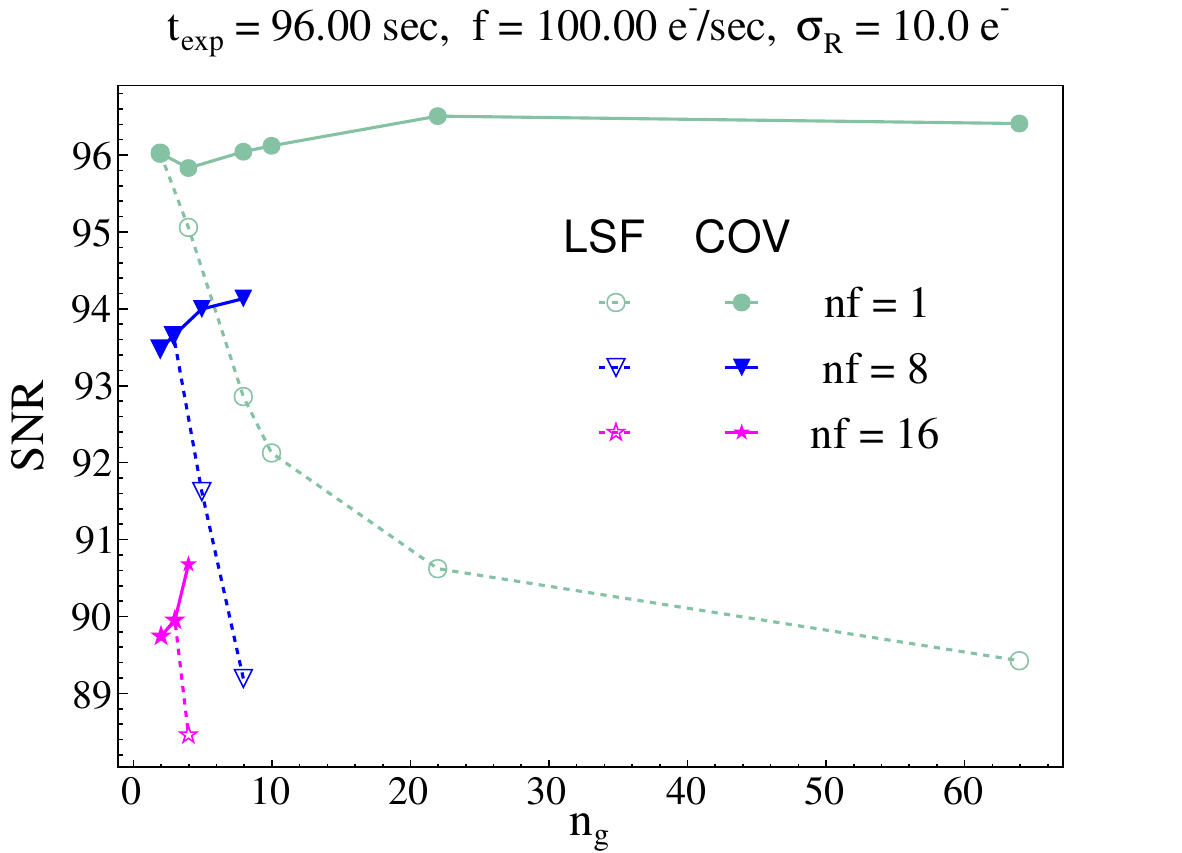}
            \includegraphics[scale=0.4]{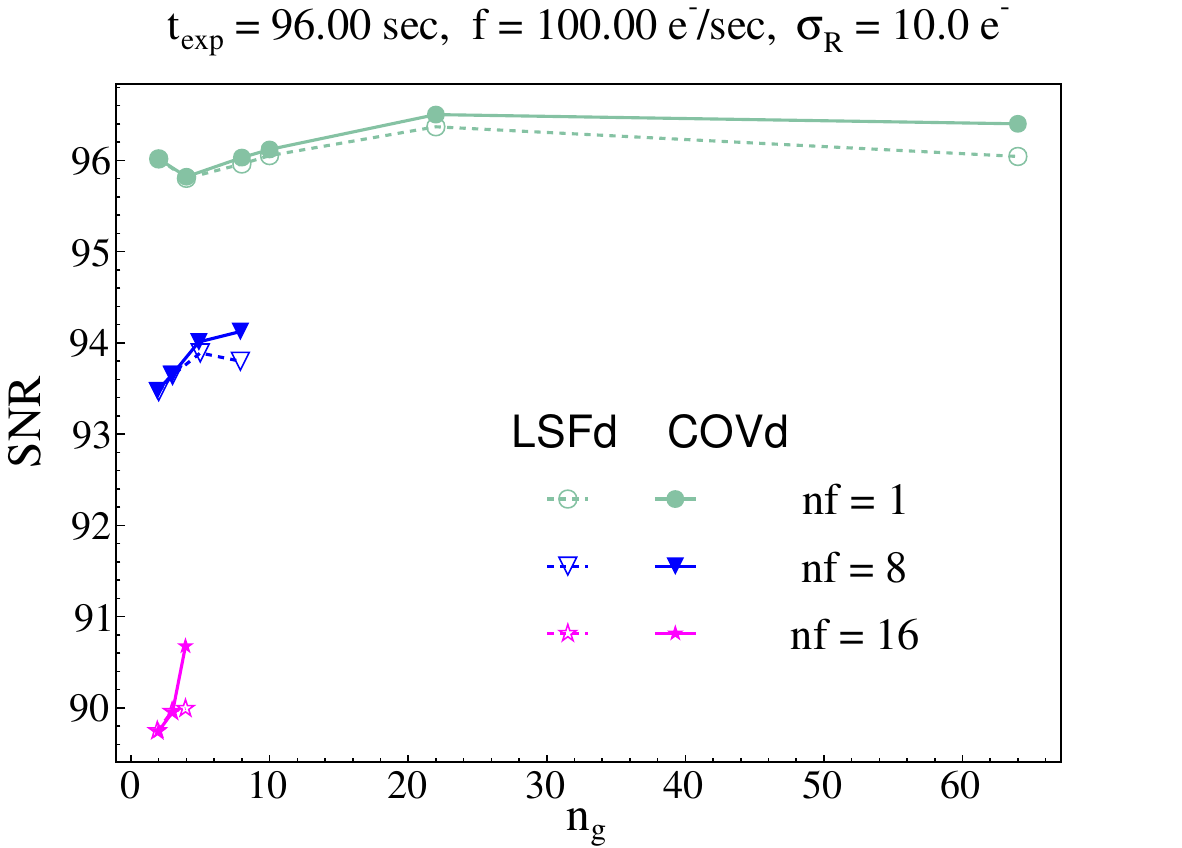}\\
        \end{tabular}
    \end{center}
    \caption[Optimal readout modes in photometric exposures]{\label{fig:photo_exposures_fits} Signal to noise ratio for photometric exposures of 96 sec exposure time in three different signal regimes (readout noise limited regime $f/\sigma_R = 0.001$, intermediate regime $f/\sigma_R = 0.1$ and background limited regime $f/\sigma_R = 10$). On the left the ramps are fitted using COV (plain line) and LSF (dotted line) fits, on the right the COVd (plain line) and LSFd (dotted line) fits are applied. In the low signal limit the best performances in terms of SNR are achieved using maximum number of frames and for all the methods of fits except LSFd the same SNR ratio of about 0.22 is reached. In the intermediate regime the differences between the different methods of fits begin to manifest. Both, COV and COVd fits reach similar SNR ratios which are maximized for maximal $n_{tot}$ as it was in low signal limit. In background limited regime the privileged fit methods are the COV, COV and LSFd, the LSF fit is the 
worst method to reach a high SNR.}
\end{figure}

\begin{figure}[ht!]
    \begin{center}
        \begin{tabular}{cc}
            \includegraphics[scale=0.4]{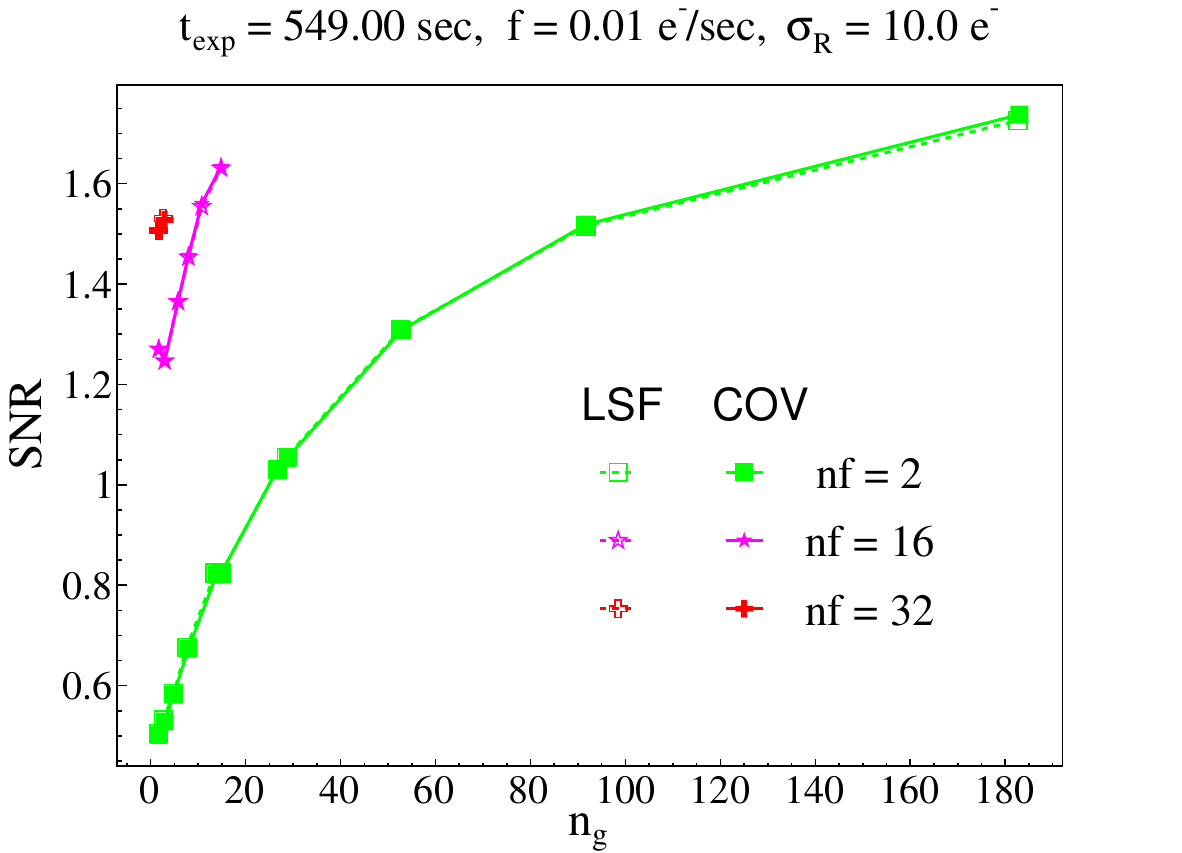}
            \includegraphics[scale=0.4]{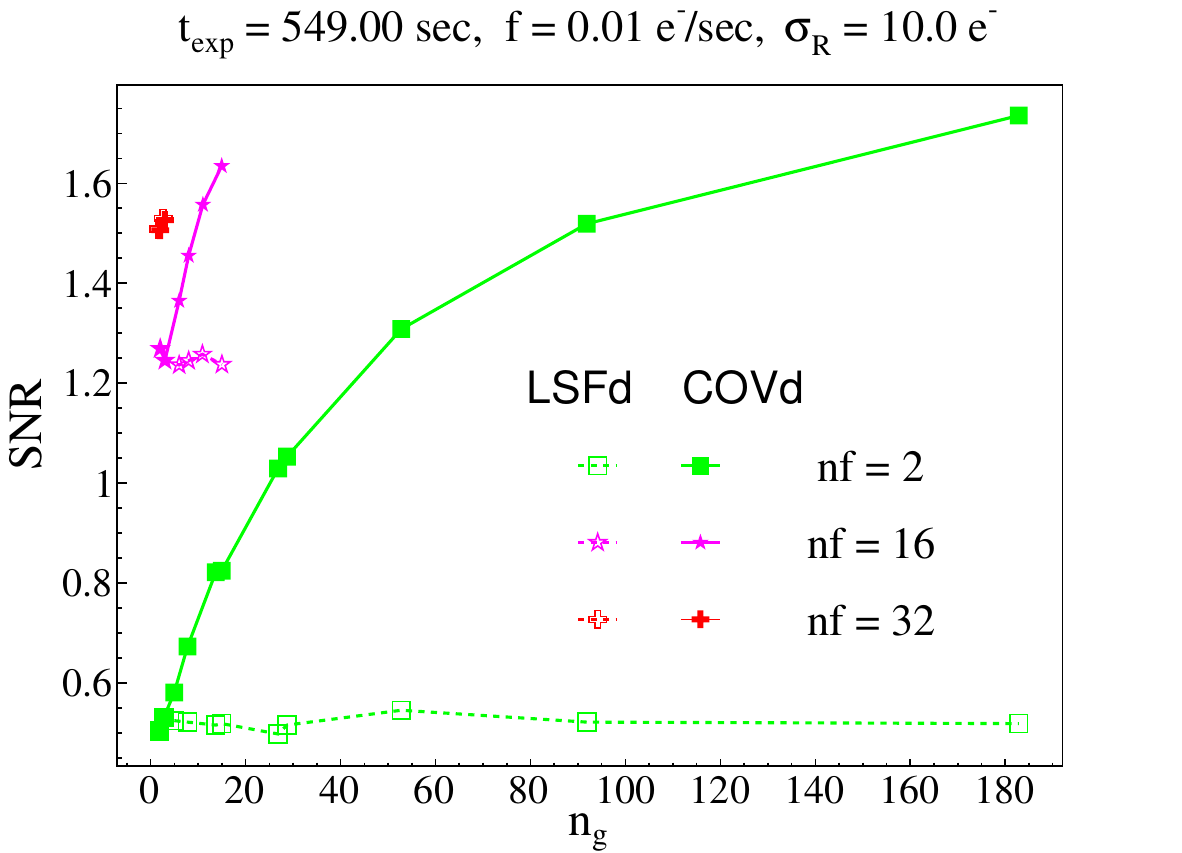}\\
            \includegraphics[scale=0.4]{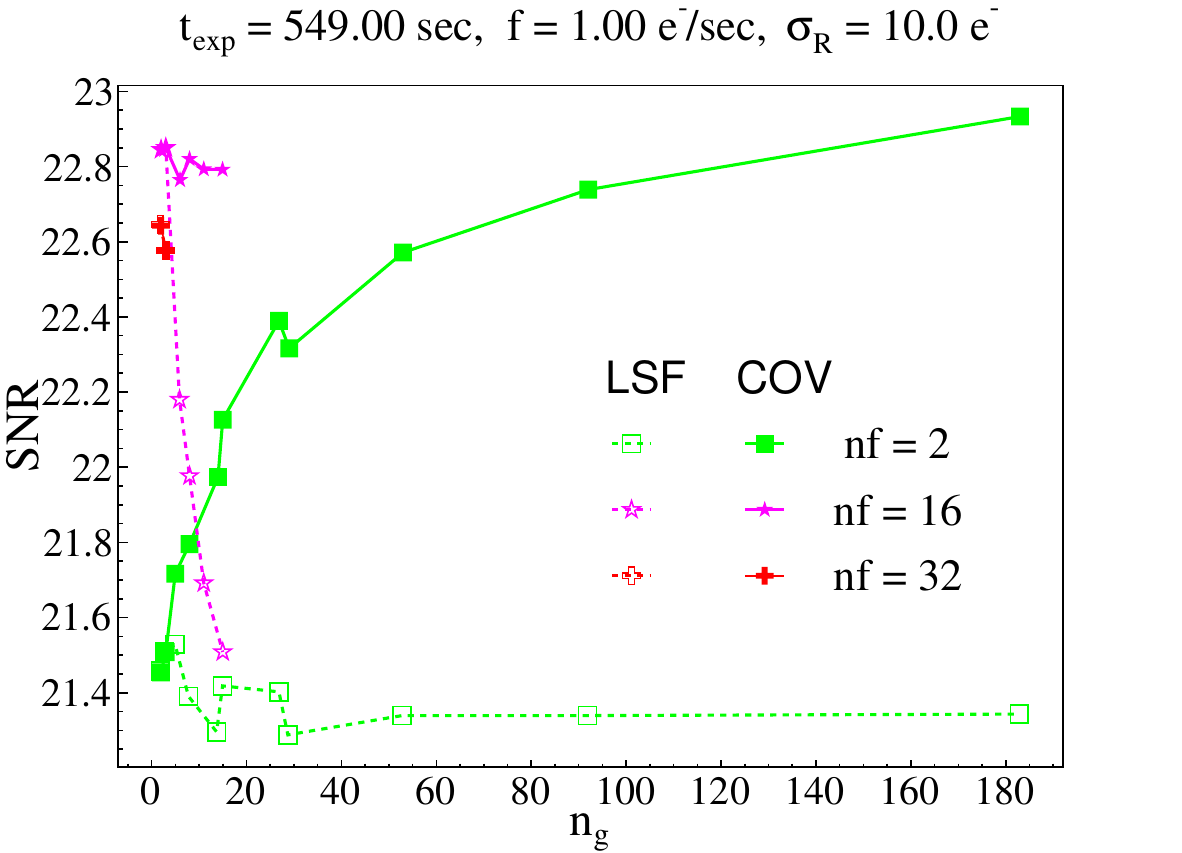}
            \includegraphics[scale=0.4]{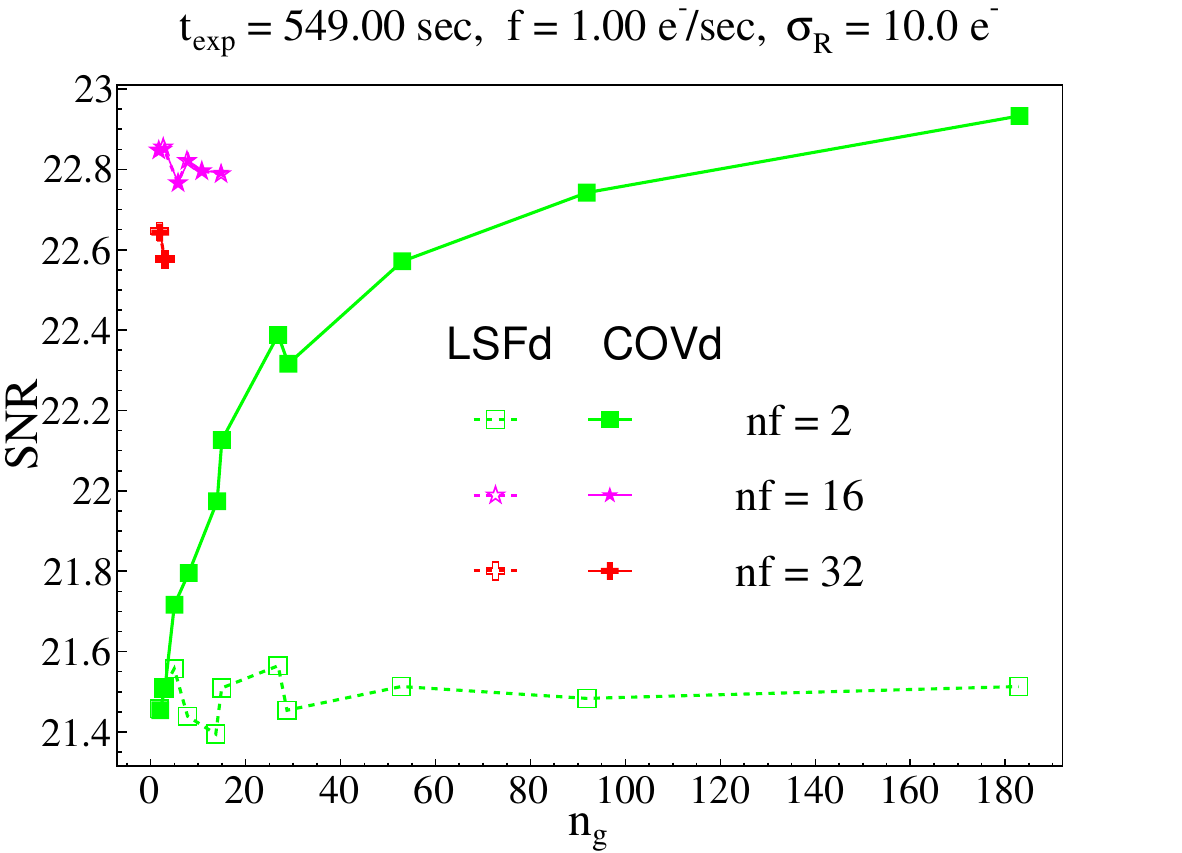}\\
            \includegraphics[scale=0.4]{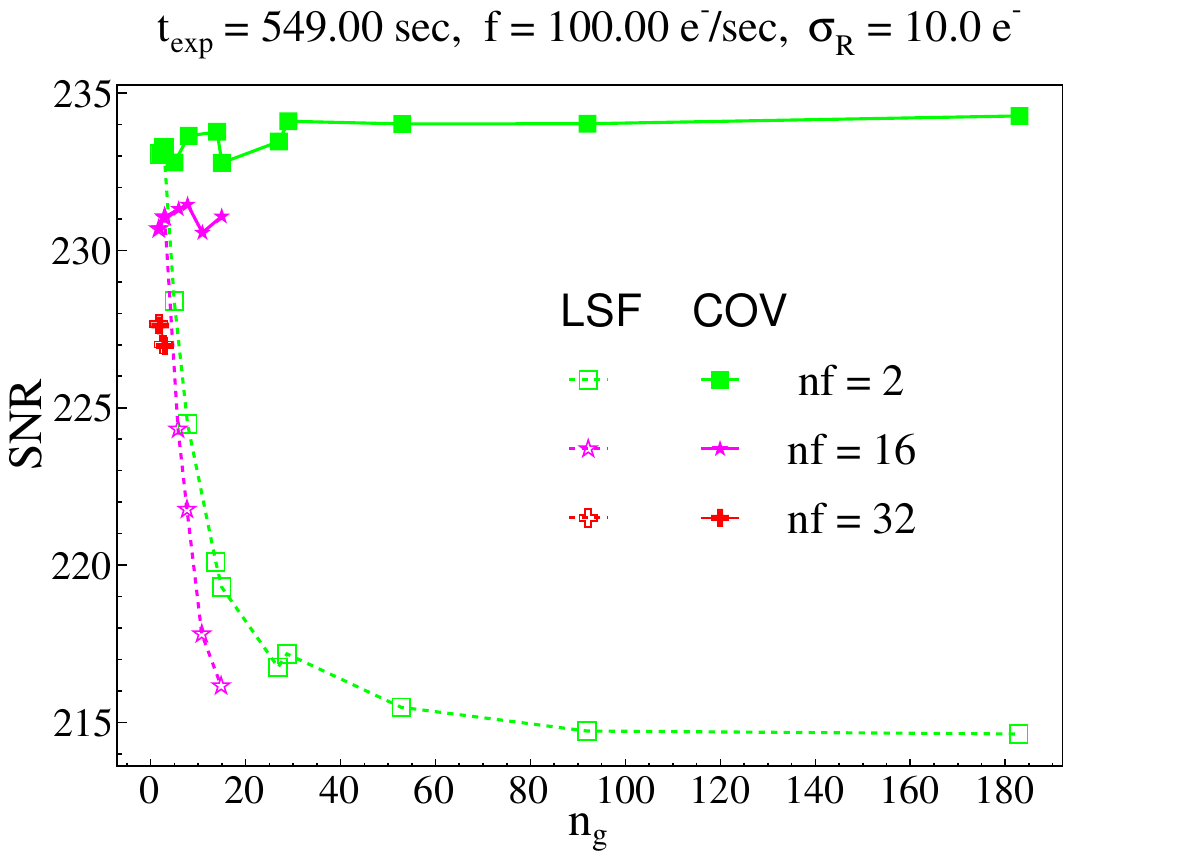}
            \includegraphics[scale=0.4]{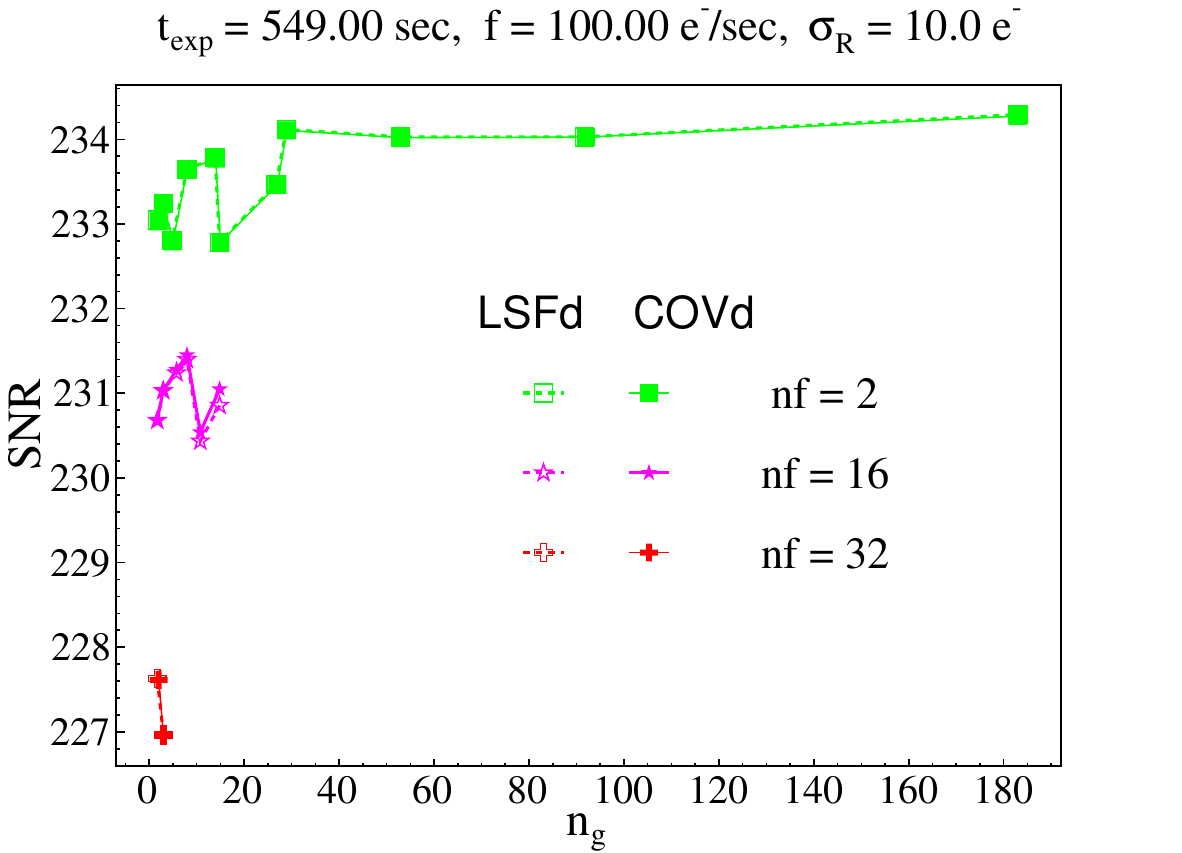}\\
        \end{tabular}
    \end{center}
    \caption[Optimal readout modes in spectroscopic exposures]{\label{fig:spectro_exposures_fits} Signal to noise ratio for spectroscopic exposures of 549 sec exposure time in three different signal regimes (readout noise limited regime $f/\sigma_R = 0.001$, intermediate regime $f/\sigma_R = 0.1$ and background limited regime $f/\sigma_R = 10$). On the left the ramps are fitted using COV (plain line) and LSF (dotted line) fits, on the right the COVd (plain line) and LSFd (dotted line) fits are applied. }
\end{figure}

We define $N$ as the total number of frames sampled up the ramp $N = n_g n_f + (n_g-1) n_d$ and $n_{tot} = n_f n_g$ as the total number of frames used effectively to fit the signal. In Figures (\ref{fig:photo_exposures_fits}) and (\ref{fig:spectro_exposures_fits}) we show the SNR ratios for photometric and spectroscopic exposures respectively for four fitting methods studied in this paper and some readout modes that could be implemented in the infrared survey missions. It is evident that in the low signal regime, where the readout noise correlation dominates, the simple up the ramp sample without coadding $n_{tot} = n_g$ should be used to maximize the signal to noise ratio with any of the COV, COVd and LSF fits. In the intermediate regime (fluxes of order of 1 $e^-$/sec), still the maximal number of frames in the ramp should be used, but the organization of frames within groups is not fixed by the SNR analysis. There is a trade-off between minimizing the readout noise of the group by averaging over many 
frames and the number of groups to trig on a sudden brake of the ramp by a spurious event such as cosmic ray. In view of cosmic ray rejection it seems to be better to maximize the number of groups $n_g$ rather than minimize the single group readout noise by coadding (too) many frames. In those two limits the LSFd fit privileges an acquisition scheme $n_g = 2$, $n_{f} = N/2$.  The situation is different in the high signal regime where the best SNR is achieved for $n_g$ = 2 in the case of LSF fit. The optimal number of frames that should be averaged depends on the ratio of the flux and the readout noise and its exact value was derived in \cite{Garnett&Forrest1993}. The fits with covariance matrices and 
the LSFd method give a possibility of arranging data into more groups with a high SNR ratio.

\newpage
\section{Conclusions}
We have evaluated analytically the covariance matrices between groups of frames arising from stochastic and readout fluctuations. The straight line fits performed with these matrices demonstrate that a significantly higher signal to noise is obtained when the fluence exceeds 1 $e^-$/sec/pix in particular in long (several hundreds of seconds) spectroscopic exposures. The improvement arising from the covariance matrix is particularly strong for the initial intercept of the fit at $t=0$, a quantity which provides a useful redundancy to cross check the signal quality. We have demonstrated that the optimal readout mode in all ranges of fluxes studied in this paper (if fit is performed using COV or COVd method) is the one that uses all the frames in the ramp. There is no restriction on the organization of frames within groups for fluxes lower than 1 $e^-$/sec/pix. For fluxes exceeding 1 $e^-$/sec/pix the coadding of frames shell be avoided. The least square fit of differences LSFd fit definitely loses information 
at low fluence as it is the case for the LSF at high fluence.

\acknowledgments
The authors would like to thank the engineers form the Institut de Physique Nucl\'eaire de Lyon that have contributed to the development of the cryogenic facility for their efforts. The authors thank also the support from the CNRS/IN2P3.
This research was conducted within the framework of the Lyon Institute of Origins under grant ANR-10-LABX-66.

\listoffigures
\listoftables

\bibliographystyle{spiebib}

\end{spacing}
\end{document}